\documentclass[a4paper,10pt]{article}

\usepackage{amssymb,amsfonts,amsmath,epsfig,float,graphicx}
\usepackage{hyperref}
\usepackage[margin=3cm]{geometry}

\newtheorem{theorem}{Theorem}[section]
\newtheorem{lemma}[theorem]{Lemma}
\newtheorem{proposition}[theorem]{Proposition}
\newtheorem{definition}[theorem]{Definition}

\newcommand{\qed}{\nobreak \ifvmode \relax \else
      \ifdim\lastskip<1.5em \hskip-\lastskip
      \hskip1.5em plus0em minus0.5em \fi \nobreak
      \vrule height0.75em width0.5em depth0.25em\fi}

\def\upd{\,{\rm d}}
\newcommand{\be}{\begin{eqnarray}}
\newcommand{\ee}{\end{eqnarray}}
\newcommand{\nnee}{\nonumber\ee}
\newcommand{\Tr}{\,{\rm Tr}\,}
\newcommand{\dom}{\,{\rm dom}\,}

\def\Co{{\mathbb C}}

\def\Io{{\mathbb I}}

\def\Mo{{\mathbb M}}

\def\Ro{{\mathbb R}}

\renewcommand{\Im}{\,{\rm Im}\,}
\renewcommand{\Re}{\,{\rm Re}\,}

\def\beginproof{\par\strut\vskip 0.0cm\noindent{\bf Proof}\par}
\def\endproof{\par\strut\hfill$\square$\par\vskip 0.2cm}

\def\Us{U^{(s)}}
\def\Usi{U^{(s)}_{i/2}}
\def\Ui{U_{i/2}}

\title{
Log-affine geodesics in the manifold of\\
vector states on a von Neumann algebra
}
\author{
Jan Naudts\\
\strut\\
\small          Departement Fysica, Universiteit Antwerpen,\\
\small          Universiteitsplein 1, 2610 Antwerpen, Belgium\\
\small          \url{Jan.Naudts@uantwerpen.be}\\
\small          \url{https://orcid.org/0000-0002-4646-1190}
}

\date {}

\begin{document}
\maketitle

\begin{abstract}
This paper introduces the notion of a log-affine geodesic connecting two vector states
on a von Neumann algebra.
The definition is linked to the standard notion of Boltzmann-Gibbs states in Statistical
Physics and the related notion of quantum statistical manifolds.
In the abelian case it is linked to the notion of exponential tangent spaces.
\end{abstract}

\section{Introduction}

The existing approach to quantum information
geometry focuses on non-degen\-erate density matrices. See for instance \cite{PD08}.
In \cite{NJ18} the present author proposes a reformulation which makes use of operators
in the commutant of the Gelfand-Naimark-Segal (GNS)
representation of the algebra of $n$-by-$n$ matrices, induced by a faithful
quantum state. The generalization of this approach to the context
of infinite-dimensional Hilbert spaces is still open.
As a first step in this direction the present paper discusses a possible
definition of log-affine geodesics in a manifold of vector states,
while leaving open all other aspects of
a full theory of statistical manifolds in a non-commutative setting.

If the geodesic is log-affine
then one expects that there exists a generator $H$ reproducing the geodesic.
Consider a von Neumann algebra $\cal A$ of operators on a Hilbert space $\cal H$,
$\Omega$ a normalized element of $\cal H$, and $T_s$ a semi-group of normal operators.
Define states $\omega_s$ on $\cal A$ by
\be
\omega_s(A)&=&(A\Omega_s,\Omega_s),
\qquad A\in{\cal A},
\nnee
with
\be
\Omega_s&=&\frac{T^{1/2}_s\Omega}{||T^{1/2}_s\Omega||},
\quad s\ge 0.
\nnee
Then $s\mapsto \omega_s$ is an example of what is meant here by a geodesic connecting
$\omega_1$ to $\omega_0$. By Stone's theorem there exists a normal operator $H$
such that $T_s=\exp(sH)$.
For this reason the geodesic is said to be log-affine.

In a statistical context it is natural to assume that the generator $H$
is a self-adjoint operator. 
A strongly continuous one-parameter
group of unitary operators $(U_t)_t$ is defined by $U_t=\exp(itH)$.
They are linked to the operator $T$ by analytic continuation: $T^{1/2}=U_{-i/2}$.

The setting as described above is too limited for the purposes of the present paper.
A more general setting is borrowed from the theory of the modular automorphism group
\cite{TM70}, which is associated with the Kubo-Martin-Schwinger (KMS) condition \cite{HHW67},
a notion coming from Statistical Physics. The proposal is to give up the requirement that 
the unitary operators $(U_t)_t$ form a group.
It is replaced by the cocycle condition \cite {MT84}
\be
U_{r+t}&=&U_r\tau_r(U_t),
\quad r,t\in\Ro,
\label{intro:cocycledef}
\ee
where $(\tau_t)_t$ is a one-parameter group of automorphisms of the
von Neumann algebra $\cal A$.
Note that in Physics the notion of a cocycle is known 
as a set of unitary operators implementing the time evolution of the interaction picture
of Quantum Mechanics.

The present work has been influenced by the approach of Pistone and Sempi \cite{PS95}
to statistical manifolds of probability distributions, and, in particular,
by the recent works of Newton \cite{NN12} and of Montrucchio and Pistone \cite{MP17,MP19} 
on the linear growth case. Part of the results of \cite{MP17} can be translated
to the non-commutative context in a rather straightforward manner \cite{NJ18arXiv}.
Translation of other parts is hindered by difficulties due to non-commutativity.
The connection with the present work is subject of future investigation.

In the next two sections the correspondence between cocycles and vector states of the von Neumann
algebra is worked out.
A definition of a log-affine geodesic is given in 
Section \ref {sect:logaff}.
Section \ref{sect:findim} shows that any two faithful quantum states defined
by $n$-by-$n$ density matrices are always connected by a log-affine geodesic.
For the abelian case, the correspondence with the notion of exponential tangent spaces
is worked out in Section \ref{sect:abel}. The final Section contains a short
summary.

\section{States labeled with cocycles} 

Throughout the text $\cal A$ denotes a von Neumann algebra which acts on a Hilbert space $\cal H$.
The symbol $\Mo$ denotes a manifold of faithful vector states
on  $\cal A$. A single state $\omega_0$ is fixed in $\Mo$.
It is generated by a cyclic and separating vector $\Omega_0\in{\cal H}$.
The one-parameter group of modular automorphisms leaving $\omega_0$
invariant is denoted $(\tau_t)_t$. The modular conjugation is denoted $S$,
the modular operator $\Delta$ euqals $S^*S$ and leaves the vector
$\Omega_0$ invariant.

Recall the following result.

\begin{lemma} [Lemma 2.1 of \cite{MT84}]
\label{logaff:lemma}
  Let be given a strongly continuous one-parameter family of unitary operators $U_t$
  belonging to $\cal A$. Assume they satisfy the cocycle condition w.r.t.~a group of 
  automorphisms $(\tau_t)_t$, which are of the form $\tau_t(A)=\Delta^{-it}A\Delta^{it}$.
  Then there exist a positive operator $T$ such that
    \be
      U_t=T^{-it}\Delta^{it},
      \qquad t\in\Ro.
    \nnee  
\end{lemma}

\beginproof
By the cocycle condition the unitary operators $V_t$, defined by
 \be
 V_t&=&U_t\Delta^{-it},
 \nnee
 satisfy
 \be
 V_{t+r}
 &=&
 U_{t+r}\Delta^{-i(t+r)}\cr
 &=&
 U_t\tau_t[U_r]\Delta^{-i(t+r)}\cr
 &=&
 U_t \Delta^{-it}U_r \Delta^{it}\Delta^{-i(t+r)}\cr
 &=&
 U_t \Delta^{-it}U_r \Delta^{-ir}\cr
 &=&
 V_tV_r.
 \nnee
 Hence $t\mapsto V_t$ is a strongly continuous one-parameter group of unitaries.
 By Stone's theorem these unitaries can be written as powers of a positive
 operator $T$. By construction one has
 \be
 T^{-it}=V_t=U_t\Delta^{-it}
 \nnee
 and hence 
 \be
 U_t&=&T^{-it}\Delta^{it}.
 \nnee
\endproof

\begin{proposition}
\label {geo:prop1}
Let be given a cocycle $(U_t)_t$ of $(\tau_t)_t$.
Let $T$ be the positive operator defined by the previous Lemma.
Are equivalent:
 \begin{description}
   \item [\quad 1) \quad]
    The map $t\mapsto U_t\Omega_0$ 
    has a continuous extension  $z\mapsto \Xi_z\in{\cal H}$
    on the strip  $0\le\Im z\le i/2$ and this extension is
    analytic  in the interior $0< \Im z < i/2$.
   \item [\quad 2) \quad]
    $\Omega_0$ belongs to the domain of $T^{1/2}$.
 \end{description}

\end{proposition}

\beginproof

\paragraph{1) $\Rightarrow$ 2)}

Use $\Delta\Omega_0=0$ to obtain
\be
U_t\Omega_0&=&T^{-it}\Omega_0.
\nnee
Let
\be
T&=&\int_0^\infty\lambda\,\upd E_\lambda
\nnee
denote the spectral representation of the operator $T$.
By assumption, the analytic continuation of
\be
F(t)=(T^{-it}\Omega_0,\psi)=\int_0^\infty\lambda^{-it}\,\upd (E_\lambda\Omega_0,\psi)
\nnee
exists for any $\psi$ in $\cal H$ up to $t=i/2$.
This implies 
\be
F(i/2)
&=&(\Xi_{i/2},\psi)\cr
&=&\int_0^\infty\lambda^{1/2}\,\upd (E_\lambda\Omega_0,\psi).
\nnee
For $\psi$ in the domain of $T^{1/2}$ one has
\be
\int_0^\infty\lambda^{1/2}\,\upd (E_\lambda\psi,\Omega_0)
&=&(T^{1/2}\psi,\Omega_0)
\nnee
and hence
\be
(T^{1/2}\psi,\Omega_0)&=&(\psi,\Xi_{i/2}).
\nnee
Because $T^{1/2}$ is self-adjoint this implies that $\Omega_0$ is in its domain
and $T^{1/2}\Omega_0=\Xi_{i/2}$.

\paragraph{2) $\Rightarrow$ 1)}
For any $z$ in the strip $0\le \Im z\le 1/2$
one has using the concavity of the logarithmic function
\be
||T^{-iz}\Omega_0||^2
&=&\int_0^\infty\lambda^{2\Im z}\,\upd (E_\lambda\Omega_0,\Omega_0)\cr
&\le&
\int_0^\infty\left[2\Im z+(1-2\Im z)\lambda\right]\,\upd (E_\lambda\Omega_0,\Omega_0)\cr
&=&2\Im z+(1-2\Im z)||T^{1/2}\Omega_0||^2.
\nnee
This upper bound suffices to prove 1) in a straightforward manner.

\endproof

\begin{proposition}
\label {geo:prop2}
Let be given a cocycle $(U_t)_t$ of $(\tau_t)_t$.
Assume that the equivalent conditions of Proposition \ref{geo:prop1} are satisfied.
Then a vector state $\omega_U$ is defined by
     \be
       \omega_U(A)=\left(A\Omega_U,\Omega_U\right),
       \qquad A\in {\cal A},
       \label {logaff:analcond}
     \ee
    with $\Omega_U=e^{-\frac 12\zeta(U)}\Xi_{i/2}=e^{-\frac 12\zeta(U)} T^{1/2}\Omega_0$
    and $\zeta(U)=\log ||\Xi_{i/2}||^2=\log ||T^{1/2}\Omega_0||^2$.

\end{proposition}

\beginproof

Let us show by proof {\em ex absurdo} that 
$\Omega_U\not=0$.
From $\Omega_U=0$ 
and $\Xi_{i/2+t}=T^{1/2+it}\Omega_0$
it follows that $\Xi_z$ vanishes along the line $\Im z=1/2$.
This implies that it vanishes everywhere on the strip $0\le \Im z\le 1/2$.
This implies that $\Xi_0=\Omega_0=0$, in contradiction with the proper
normalization of $\Omega_0$.

The remainder of the proof is straightforward.

\endproof

\section{Some operators and their properties}

\begin{proposition}
\label {geo:prop3}
Let be given a cocycle $(U_t)_t$ of $(\tau_t)_t$.
Assume that the equivalent conditions of Proposition \ref{geo:prop1} are satisfied.
Then a linear operator, denoted $\Ui$, is defined by
\be
\Ui Y\Omega_0&=&Y\Xi_{i/2}=YT^{1/2}\Omega_0,
\quad Y\in{\cal A}'.
\nnee
It has the following properties. 
 \begin{description}
   \item [\quad 1) \quad]
    For any $Y$ in the commutant ${\cal A}'$ the vector $\Ui Y\Omega_0$ is the analytic extension of 
    the map $t\mapsto U_tY\Omega_0$ to the point $t=i/2$;
   \item [\quad 2) \quad]
    $\Xi_{i/2}=\Ui \Omega_0=T^{1/2}\Omega_0$;
   \item [\quad 3) \quad]
    $\Ui $ commutes with any operator in the commutant ${\cal A}'$.
 \end{description}

\end{proposition}

\beginproof
Because $\Omega_0$ is cyclic for the commutant the linear operator $\Ui $
is a densily defined operator.

For any $Y,Z$ in the commutant ${\cal A}'$
the function $t\mapsto(U_tY\Omega_0,Z\Omega_0)$ has an analytic extension
to $t=i/2$. Indeed, one has
\be
(U_tY\Omega_0,Z\Omega_0)&=&(U_t\Omega_0,Y^*Z\Omega_0).
\nnee
By assumption the r.h.s.~has an analytic extension up to $t=i/2$ where it has the value
$(\Xi_{i/2},Y^*Z\Omega_0)$.
Therefore one has
\be
(U_tY\Omega_0,Z\Omega_0)\bigg|_{t=i}
&=&
(\Xi_{i/2},Y^*Z\Omega_0)\cr
&=&
(YT^{1/2}\Omega_0,Z\Omega_0)\cr
&=&
(\Ui Y\Omega_0,Z\Omega_0).
\nnee
This shows that $\Ui Y\Omega_0$ is the analytic extension of $U_tY\Omega_0$
to $t=i/2$. The argument can also be used to show that $\Ui $ commutes
with any element of the commutant.

\endproof

\begin{proposition}
\label {geo:prop4}
Let be given a cocycle $(U_t)_t$ of $(\tau_t)_t$.
Assume that the equivalent conditions of Proposition \ref{geo:prop1} are satisfied.
Assume in addition that $\Delta$ is the modular operator.
Let $\Ui $ be the operator defined in the previous Proposition.
Let $S=J\Delta^{1/2}$ denote the polar decomposition of the modular conjugation operator.
The following two conditions are equivalent.
\begin{description}
  \item [\quad 1) \quad] $\Xi_{i/2}$ is in the domain of $S$.

  \item [\quad 2) \quad] $\Omega_0$ belongs to the domain of $\Ui^*$.

\end{description}
If these conditions are satisfied then the following hold.
\begin{description}
  \item [\quad 3) \quad] The operator $\Ui $ is closable.

  \item [\quad 4) \quad] $\Ui^*\Omega_0=S\Xi_{i/2}$.

  \item [\quad 5) \quad] $J\Xi_{i/2}=\Xi_{i/2}$.
  
  \item [\quad 6) \quad] The operator $Y=J\Ui^{**}J$ is affiliated with the commutant ${\cal A}'$;
   The operator $JT^{1/2}S$ is an extension of the operator $Y$; 
   In particular, one has $Y\Omega_0=\Xi_{i/2}$.

  \item [\quad 7) \quad] The state $\omega_U$, defined in Proposition \ref {geo:prop2}, satisfies
    \be
      \omega_U(A)
      =e^{-\zeta(U)}(AY\Omega_0,Y\Omega_0)
      =e^{-\zeta(U)}(AX^{1/2}\Omega_0,X^{1/2}\Omega_0),
      \qquad A\in {\cal A},
    \nnee
    with $X=Y^*Y$.

\end{description}

\end{proposition}

\beginproof

\paragraph{1) implies 2)}
One has for all $Y$ in the commutant ${\cal A}'$ that
\be
(S\Xi_{i/2}, Y\Omega_0)
&=&
(S^*Y\Omega_0,\Xi_{i/2})\cr
&=&
(Y^*\Omega_0,\Xi_{i/2})\cr
&=&
(\Omega_0,Y\Xi_{i/2})\cr
&=&
(\Omega_0,\Ui Y\Omega_0).
\nnee
This shows that $\Omega_0$ is in the domain of $\Ui^*$
and that $\Ui^*\Omega_0=S\Xi_{i/2}$.

\paragraph{2) implies 1) and 4)}
For any $Y$ in the commutant ${\cal A}'$ one has
\be
(\Xi_{i/2},S^*Y\Omega_0)
&=&
(\Ui \Omega_0,Y^*\Omega_0)\cr
&=&
(Y\Omega_0,\Ui^*\Omega_0).
\nnee
This implies that $\Xi_{i/2}$ is in the domain of $S$ and that 
$S\Xi_{i/2}=\Ui^*\Omega_0$.

\paragraph{2) implies 3)}
Assume that $Y_n\Omega_0$ converge to 0 and $\Ui Y_n\Omega_0$
converge to some $\psi$ in $\cal H$.
Then one has for all $Y,Z$ in the commutant that
\be
(\psi,Z\Omega_0)
&=&\lim_n(\Ui Y_n\Omega_0,Z\Omega_0)\cr
&=&\lim_n(\Ui Z^*Y_n\Omega_0,\Omega_0)\cr
&=&\lim_n(Z^*Y_n\Omega_0,\Ui^*\Omega_0)\cr
&=&0.
\nnee
One concludes that $\psi=0$. Hence, the operator $\Ui$ is closable.

\paragraph{5)}
In the case of a finite-dimensional Hilbert space $\cal H$ the proof is simple.
From $\Ui=T^{1/2}\Delta^{-1/2}$ it follows that $\Ui^*=\Delta^{-1/2}T^{1/2}$.
One then calculates, using that $\Ui$ belongs to $\cal A$,
\be
\Xi_{i/2}
&=&
\Ui\Omega_0\cr
&=&
S\Ui^*\Omega_0\cr
&=&
S\Delta^{-1/2}T^{1/2}\Omega_0\cr
&=&
J\Xi_{i/2}.
\nnee

The general proof is found in the Appendix.

\paragraph{6)}
Because $\Ui^{**}$ is affiliated with $\cal A$
the operator $Y=J\Ui^{**}J$ is affiliated with the commutant ${\cal A}'$.

In the case of a finite-dimensional Hilbert space $\cal H$ the proof 
of the equality $Y=JT^{1/2}S$ is simple.
From $\Ui=T^{1/2}\Delta^{-1/2}$
it follows that
\be
Y&=&J\Ui J=JT^{1/2}\Delta^{-1/2}J=JT^{1/2}S.
\nnee
The general proof is found in the Appendix.

\paragraph{7)}
The polar decomposition of the operator $Y$ can be written as $Y=KX^{1/2}$.
Because the isometry $K$ belongs to ${\cal A}'$ one has for all $A$ in $\cal A$
that
\be
\omega_U(A)&=&e^{-\zeta(U)}(A\Xi_{i/2},\Xi_{i/2})\cr
&=&e^{-\zeta(U)}(AY\Omega_0,Y\Omega_0)\cr
&=&e^{-\zeta(U)}(AKX^{1/2}\Omega_0,KX^{1/2}\Omega_0)\cr
&=&e^{-\zeta(U)}(AX^{1/2}\Omega_0,X^{1/2}\Omega_0).
\nnee

\endproof

\section{Log-affine geodesics}
\label{sect:logaff}
\def\LL{L^{\mbox{\tiny L}}}
\def\LR{L^{\mbox{\tiny R}}}

\def\Hi{H^{\mbox{\tiny L}}}

Let $\cal A$, $\Mo$, $\omega_0$ and $\Omega_0$ be as before.
The state $\omega_0$ is invariant for the one-parameter group
$(\tau_t)_t$ of automorphisms of $\cal A$.
These automorphisms are of the form $\tau_t(A)=\Delta^{-it}A\Delta^{it}$,
where $\Delta$ is a positive operator satisfying $\Delta\Omega_0=0$.

Let us now consider a geodesic $s\in[0,1]\mapsto\omega_s$ which connects
two states $\omega_0$ and $\omega_1$ in $\Mo$. It is assumed that
each state $\omega_s$ is labeled by a cocycle $U^{(s)}$ which satisfies
the equivalent conditions of Proposition \ref {geo:prop4}.
The state $\omega_{\Us}$ defined in Proposition \ref {geo:prop2}
coincides with the state $\omega_s$.
The normalization function is denoted $\zeta(s)$ instead of $\zeta(\Us)$.
The operator $T$ constructed in Lemma \ref {logaff:lemma} is denoted $T_s$.

\subsection{Logarithmic derivatives}

Logarithmic derivatives play an essential role in Quantum Information Theory.
Consider an operator-valued
function $s\mapsto T_s\equiv\exp(H_s)$. In general, the derivative $\upd H_s/\upd s$ does not
commute with the operator $T_s$. As a consequence, the derivative 
$\upd T/\upd s$ can differ from 
\be
T_s\frac {\upd H}{\upd s}\,\mbox{ or }\frac {\upd H}{\upd s}T_s.
\nnee
It would be fair to call $\upd H_s/\upd s$ the logarithmic derivative of $s\mapsto T_s$.
However, this does not help us in the calculation of $\upd T/\upd s$.
Following \cite{PT93} let us call the {\em left logarithmic derivative} of $s\mapsto T_s$
any solution of the problem
\be
\frac{\upd T}{\upd s}&=&\LL_sT_s.
\nnee
Similarly, the {\em right logarithmic derivative} $\LR_s$ is any solution of the problem
\be
\frac{\upd T}{\upd s}&=&T_s\LR_s.
\nnee
On the other hand, the {\em symmetric logarithmic derivative} is
the solution of
\be
\frac{\upd T}{\upd s}&=&\frac 12(L_sT_s+T_sL_s).
\nnee
See for instance \cite {LCJW16} and the references quoted there.
If $\LL_s=\LR_s$ then they coincide with the symmetric logarithmic derivative $L_s$

\begin{proposition}
 Assume the following.
 \begin{description}
  \item [\quad -\quad]
     The equivalent conditions of Proposition \ref {geo:prop4} are satisfied. 
  \item [\quad -\quad]
     There exists an operator $\Hi_s$ which solves the equation
    \be
      \frac{\upd\,}{\upd s}T^{1/2}_s
      &=&\frac 12\Hi_sT^{1/2}_s
      \label{deriv:logder}
    \ee
    on a domain which includes $\Omega_0$. 
  \item [\quad -\quad]
    The normalization function $\zeta(s)$ is differentiable.
 \end{description}
 Then one has
 \begin{description}
 
  \item [\quad 1) \quad] $\displaystyle
    \frac{\upd\,}{\upd s}\Omega_s
    =
    \left[
    \Hi_s-\frac 12\zeta'(s)
    \right]\Omega_s$.

  \item [\quad 2) \quad] $\displaystyle
    \frac{\upd\,}{\upd s}Y_s=(J\Hi_sJ)Y_s$ holds on a domain which includes $\Omega_0$.

  \item [\quad 3) \quad] An element $f_s$ of the dual of $\cal A$ is defined by
  \be
    f_s(A)&=&\left(\frac 12\left[J\Hi_sJ A+AJ\Hi_sJ\right]\Omega_s,\Omega_s\right)
    -\omega_s(A)\frac{\upd\zeta}{\upd s},
    \quad A\in{\cal A};
  \nnee
  It satisfies $f_s(\Io)=0$.
  
 \item [\quad 4) \quad]
   \be
     \frac{\upd\,}{\upd s}\omega_s(A)&=&f_s(A)
     \quad\mbox{ for all }A\in{\cal A}.
     \label{prop:tangent}
   \ee
 \end{description}

\end{proposition}

\beginproof

\paragraph{1)}
One calculates
\be
\frac{\upd\,}{\upd s}\Omega_s
&=&
\frac{\upd\,}{\upd s}e^{-\frac 12\zeta(s)}T_s^{1/2}\Omega_0\cr
&=&
-\frac 12\zeta(s)\Omega_s+e^{-\frac 12\zeta(s)}\Hi_sT_s^{1/2}\Omega_0\cr
&=&
\left[
   \Hi_s-\frac 12\zeta'(s)
   \right]\Omega_s.
\nnee

\paragraph{2)}
This follows immediately from $Y_s\subset JT^{1/2}S$, proved in Proposition \ref{geo:prop4}.

\paragraph{3) and 4)}
The proof is straightforward.

\endproof

The functional $f_s$ is a {\em tangent vector}, tangent to the geodesic $s\mapsto\omega_s$
at the point $s$. The assumptions made in the above Proposition are technical requirements
which guarantee the existence of tangent vectors. They do not suffice to say that
the geodesic is log-affine. A proposal to define the latter follows below.

\subsection{Definition of a log-affine geodesic}
\label{subsect:def}

The operator $T_s$ is uniquely determined by the cocycle $(\Us_t)_t$. This justifies
the introduction of the following definition.

\begin{definition}
\label {def:logaff}
   Let be given a map $s\in[0,1]\mapsto \omega_s\in\Mo$.
   Assume that each state $\omega_s$ is labeled with a strongly continuous
   map $t\mapsto \Us_t$, which forms a cocycle for the automorphism group
   $(\tau_t)_t$, in such a way that
   \be
     \omega_s(A)&=&e^{-\zeta(s)}(A\Usi  \Omega_0,\Usi \Omega_0),
     \qquad A\in{\cal A}.
     \nnee
   Let $T_s$ denote the corresponding positive operator,
   as defined in Lemma \ref {logaff:lemma}.
   The map $s\in[0,1]\mapsto \omega_s\in\Mo$ is said to be a {\em log-affine geodesic
   connecting $\omega_1$ to $\omega_0$} if 
   $U^{(0)}_t=\Io$ for all $t$
   and there exist a self-adjoint operator $H$
   and a function $\Phi(s)$ such that for all $s,r\in [0,1]$
   one has 
   \be
     [\log T_s+\Phi(s)]-[\log T_r+\Phi(r)]&=&(s-r)H.
    \label{logaff:logaffdef}
   \ee

\end{definition}

\vskip 8pt

From $U^{(0)}_t=\Io$ for all $t$ it follows that $T_0=\Delta$.
From (\ref {logaff:logaffdef})
it follows that the derivative exists and is given by
    \be
       \frac{\upd\,}{\upd s}[\log T_s+\Phi(s)]\psi&=&H\psi,
       \quad \psi\in\dom H.
       \label{logaff:logaffderiv}
    \ee
The cocycles of a log-affine geodesic can be written as
\be
\Us_t&=&e^{it(\Phi(s)-\Phi(0)}e^{-it[\log \Delta-sH]}\Delta^{it}.
\nnee

In the example of a semigroup, discussed in the Introduction,
the operators $\Us_t$ and $T_s$ are of the form $\Us_t=e^{istH}$, respectively
$T_s=e^{sH}$.
This corresponds with $\Delta=\Io$ and $\Phi(s)$ a constant function.
Hence, one has $\Usi=T^{1/2}_s$ with $T_s=\exp[-sH]$.
The state $\omega_0$ is then a tracial state.
This shows that (\ref {logaff:logaffdef}) is a straightforward generalization
from a tracial state together with a semigroup to a faithful vector state
together with a family of cocycles.

A further justification for the above definition comes from the compatibility with
the existing notions of quantum exponential families and of
exponential tangent spaces. See Sections \ref {sect:findim}, respectively \ref{sect:abel}.

It is straightforward to verify that
if $s\mapsto\omega_s$ is a log-affine
geodesic connecting $\omega_1$ to $\omega_0$ then for any $\lambda \in [0,1]$
the map $s\mapsto\omega_{\lambda s}$ is a log-affine
geodesic connecting $\omega_\lambda$ to $\omega_0$.

\section{The matrix algebra}
\label{sect:findim}

Let $H_1,H_2,\cdots H_k$ be self-adjoint $n$-by-$n$ matrices.
They define a density matrix $\rho_\theta$ by
\be
\rho_\theta&=&\frac{1}{Z(\theta)}\exp(-\sum_{j=1}^k\theta^j H_j).
\label{ex:rhodef}
\ee
Here, $\theta$ belongs to some open convex domain $D\subset\Ro^k$.
The partition sum $Z(\theta)$ is given by
\be
Z(\theta)&=&\Tr \exp\left(-\sum_{j=1}^k\theta^j H_j\right).
\nnee
The expression (\ref {ex:rhodef}) is the quantum analogue of a Boltzmann-Gibbs distribution.
The operators $H_1,H_2,\cdots H_k$ have the meaning of parts of a Hamiltonian.

A state $\omega_\theta$ on the algebra $\cal A$ of all $n$-by-$n$ matrices
is defined by
\be
\omega_\theta(A)&=&\Tr\rho_\theta A,\qquad A\in {\cal A}.
\nnee
By the Gelfand-Naimark-Segal (GNS) construction the elements $A$ of $\cal A$
act on a Hilbert space $\cal H$ in which there exists a cyclic and separating vector
$\Omega_\theta$ such that
\be
\Tr \rho_\theta A&=&(A\Omega_\theta,\Omega_\theta)
\quad\mbox{ for all } A\mbox{ in }{\cal A}.
\nnee

Fix yet another set of parameters $\eta$ in $D$ and let
$H=-\sum_{j=1}^k(\eta^j-\theta^j) H_j$.
Use the abbreviation $\omega_s=\omega_{(1-s)\theta+s\eta}$.
The corresponding density matrix $\rho_{(1-s)\theta+s\eta}$ is denoted $\rho_s$.

Let $\Delta$ denote the modular operator corresponding to the state $\omega_\rho$
and select $\Omega_0$ such that $\Delta\Omega_0=\Omega_0$.
The modular automorphisms $\tau_t$ satisfy
\be
\tau_t[A]=\Delta^{-it}A\Delta^{it}=\rho_0^{-it}A\rho_0^{it}
\qquad A\in{\cal A}.
\nnee
Let $\Us_t=\rho_s^{-it}\rho_0^{it}$. It is the
product of two unitary operators and hence it is itself 
a unitary operator as well. It belongs to $\cal A$.
One verifies that the cocycle condition is satisfied:
\be
\Us_{r+t}&=&\rho_s^{-i(r+t)}\rho_0^{i(r+t)}\cr
&=&
\Us_r\rho_0^{-ir}\rho_s^{-it}\rho_0^{i(r+t)}\cr
&=&
\Us_r\tau_r[\Us_t].
\nnee

Let $\Xi^{(s)}_z=\rho_s^{-iz}\rho_0^{iz}\Omega_0$.
Then one has $\Xi^{(s)}_{i/2}=\rho_s^{1/2}\rho_0^{-1/2}\Omega_0$ so that
for all $A\in{\cal A}$
\be
\left(A\Xi^{(s)}_{i/2},\Xi^{(s)}_{i/2}\right)
&=&
\left(A\rho_s^{1/2}\rho_0^{-1/2}\Omega_0,\rho_s^{1/2}\rho_0^{-1/2}\Omega_0\right)\cr
&=&
\Tr\rho_0\left(\rho_0^{-1/2}\rho_s^{1/2}A\rho_s^{1/2}\rho_0^{-1/2}\right)\cr
&=&
\Tr \rho_s A\cr
&=&\omega_s(A).
\nnee
Hence, (\ref{logaff:analcond}) is satisfied with $\zeta(s)=1$ for all $s$.

Note that the operator $\Delta\rho^{-1}_0$ belongs to the commutant ${\cal A}'$.
This follows from
\be
(\Delta\rho_0^{-1}AB\Omega_0,C\Omega_0)
=
(C^*\Omega_0,B^*A^*\rho_0^{-1}\Omega_0)=\Tr ABC^*=\Tr BC^*A
=(\Delta\rho_0^{-1}B\Omega_0,A^*C\Omega_0).
\nnee
Hence $T_s=\Delta\rho_0^{-1}\rho_s$
implies that
\be
\log T_s
&=&
\log \rho_s-\log \rho_0\Delta^{-1}
\nnee
and
\be
\frac{\upd\,}{\upd s}[\log T_s+\Phi(s)]
&=&
H,
\nnee
with
\be
\Phi(s)&=&\log Z(s)\equiv\log Z\left((1-s)\theta+s\eta\right).
\nnee
One also has
\be
\log T_s-\log T_0
&=&
sH-\Phi(s)+\Phi(0).
\nnee
On the other hand, the left logarithmic derivative is given by
\be
\frac 12\Hi_s&=&\left(\frac{\upd\,}{\upd s}T_s^{1/2}\right)T_s^{-1/2}\cr
&=&
\frac 12\int_0^1\upd u\,T_s^{u/2}\left(H-\Phi'(s)\right)T_s^{(1-u)/2}.
\nnee
It is self-adjoint and therefore it coincides with the right logarithmic
derivative and the symmetric logarithmic derivative.

In \cite{NJ18} an operator $X_s$ is introduced which belongs to the commutant
${\cal A}'$ and is defined by $X_s\Omega_0=\rho_s\rho_0^{-1}\Omega_0$,
in the notations of the present paper. A short calculation shows that
this operator satisfies $X_s=\rho_0^{-1}S^*\rho_s S=S^*T_s S$ 
and hence that it coincides with the
operator $X_s=Y_s^*Y_s$ as introduced in Proposition \ref {geo:prop4}.

\section{The abelian case}
\label{sect:abel}

In this section the von Neumann algebra $\cal A$ is (isomorphic with) the space
$L_\infty(\Ro^n,\Co)$ of all essentially bounded
complex functions on $\Ro^n$ with its Lebes\-gue measure. The Hilbert space $\cal H$
coincides with the space of square-integrable complex functions.
A function $f(x)$, element of $\cal A$,
acts on a square-integrable function $g(x)$ by pointwise multiplication $(fg)(x)=f(x)g(x)$.
The commutant ${\cal A}'$ of $\cal A$ coincides with $\cal A$.
See \cite {DJ81}, Part I, Ch. 7, Thm. 2.

Consider states $\omega_\psi$ of $\cal A$ of the form
\be
\omega_\psi(f)&=&\int_{\Ro^n}f(x)\,|\psi(x)|^2\,\upd x,
\nnee
where $\psi\in{\cal H}$ is a square-integrable function with $L_2$ norm
\be
||\psi||_2&=&\left[\int_{\Ro^n}|\psi(x)|^2\,\upd x\right]^{1/2}
\nnee
equal to 1.
The states $\omega_\psi$ can be identified with probability measures on $\Ro^n$.
Of interest is the set $\Mo$ of states $\omega_\psi$ where
$|\psi(x)|^2$ is non-vanishing almost everywhere. This implies that
the state $\omega_\psi$ is faithful and that the vector $\psi$ is cyclic
and separating for the von Neumann algebra $\cal A$.

The modular operator $\Delta$ is the identity, i.e.~the constant function 1.
Therefore, any cocycle $U(t)$ is of the form $U(t)=\exp(ith)$ with $h$
a measurable function. The analytic continuation $\Xi_z$
of $U(t)\psi$ to $t=i/2$ should belong to $\cal H$.
This is,
\be
\Xi_{i/2}(x)&=&e^{-\frac 12h(x)}\psi(x)
\nnee
must be square integrable.
If this is the case, then a state $\omega_h$ is defined by
\be
\omega_h(f)=e^{-\zeta(h)}(f\Xi_{i/2},\Xi_{i/2})
=e^{-\zeta(h)}\int\upd x\,f(x)e^{-\Re h(x)}|\psi(x)|^2.
\nnee
The normalization $\zeta(h)$ is given by
\be
\zeta(h)&=&\log \int\upd x\,e^{-\Re h(x)}|\psi(x)|^2\cr
&=&\omega_\psi(e^{-\Re h(x)}).
\nnee

\begin{definition} [Definition 3.14 of \cite{AJVS18}]
Fix a single $\omega_\psi$. A measurable real function $k(x)$ is said
to belong to the {\em  exponential tangent space} at the point $\omega_\psi$
if it satisfies the condition
\be
\int\left[e^{|tk|}-1\right]\,|\psi(x)|^2\,\upd x<+\infty
\quad\mbox{for some } t\in\Ro, t\not=0.
\label{abel:cond}
\ee
\end{definition}

\begin{lemma}
\label {lemma:abe1}
Assume that the real function $k(x)$ belongs to the exponential space tangent at $\omega_\psi$.
Then the integrals
 \be
 \int e^{\pm tk(x)}|\psi(x)|^2\,\upd x
 \label{abe:converge}
 \ee
 converge for any $t$ for which (\ref {abel:cond}) holds.
 \end{lemma}
 
\beginproof
Split the function $k(x)$ into its positive and its negative part: $k=k_+-k_-$.
Then (\ref {abel:cond}) can be written as
\be
\int e^{tk_+}\,|\psi(x)|^2\,\upd x <+\infty
\quad\mbox{ and }\quad 
\int e^{tk_-}\,|\psi(x)|^2\,\upd x <+\infty.
\nnee
Since the exponential function is
bounded for negative values of the argument and $\psi(x)$ is square-integrable
one concludes that (\ref {abe:converge}) holds.

\endproof

 \begin{lemma}
 \label {lemma:abe2}
Assume that $h(x)$ is locally square-integrable.
Then the operator $H$, defined by $(H\phi)(x)=h(x)\phi(x)$,
is a densely-defined self-adjoint operator on $\cal H$ affiliated with $\cal A$.
\end{lemma}

\beginproof
The continuous functions $\phi$ vanishing outside a compact subset $C_\phi$ of $\Ro^n$
form a dense subspace of $\cal H$ and belong to the domain of $H$ because
\be
\int |H\phi(x)|^2\,\upd x&\le&||\phi||^2_\infty \int_{C_\phi} |h(x)|^2\,\upd x
\,<\,+\infty.
\nnee
Hence the operator $H$ is densely defined. It is self-adjoint on the domain
\be
\left\{\phi:\,\int h^2(x)\,|\phi(x)|^2\,\upd x<+\infty\right\}.
\nnee.

\endproof

\begin{proposition}
Fix a normalized square-integrable function $\psi(x)$ which is non-vanishing almost everywhere.
Let be given a real function $k(x)$ which belongs to the exponential space tangent at $\omega_\psi$.
Assume in addition that $k(x)$ is locally square-integrable.
Select $t$ for which (\ref {abe:converge}) holds and let $h(x)=tk(x)$.
Let $H$ be the self-adjoint operator discussed in Lemma \ref{lemma:abe2}.
Then one has
\begin{description}
\item [1)] $\psi$ belongs to the domain of the  operators $\exp(\pm H/2)$; 
\item [2)] Let $\Omega_1$ be given by
 \be
 \Omega_1=[e^{H-\zeta(1)}]^{1/2}\psi,
 \quad\mbox{ with }\quad \zeta(1)=\log ||e^{H/2}\psi||^2.
 \nnee
Then the state $\omega_{1}$ defined by the vector $\Omega_1$
is connected to the state $\omega_\psi$ by a log-affine geodesic.
\end{description}

\end{proposition}

\beginproof

\paragraph{1)}

This follows immediately from Lemma \ref{lemma:abe1}.

\paragraph{2)}
Let
\be
H&=&\int\lambda\upd E_\lambda
\nnee
be the spectral decomposition of $H$.
Let $0\le s\le 1$ and $\beta\in\Ro$.
Convexity of the exponential function implies that
\be
\int \exp(s\lambda-\beta)\,\upd(E_\lambda\psi,\psi)
&\le&
e^{-\beta}\left[1-s+s\int \exp(\lambda)\,\upd(E_\lambda\psi,\psi)\right].
\nnee
The latter integral is finite because $\psi$ belongs to the domain of $\exp H/2$.
One concludes from the inequality that $\psi$ belongs to the domain of
$[\exp(s\lambda-\beta)]^{1/2}$ for all real $\beta$.

Let
\be
\omega_s(A)&=&(A\Omega_s,\Omega_s)
\quad\mbox{ for all }A\in{\cal A},
\nnee
with
\be
\Omega_s=[\exp(sH-\zeta(s))]^{1/2}\psi
\quad\mbox {and}\quad
\zeta(s)=\log||\exp(sH/2)\psi||.
\nnee
Note that $\Omega_0=\psi$ and $\omega_0=\omega_\psi$.

Let us now verify that $s\mapsto\omega_s$ is a log-affine connection of $\omega_1$
with $\omega_0$. In the present commutative context the one-parameter group of
automorphisms $(\tau_t)_t$ is the trivial one. The modular operator $\Delta$
is the identity operator. The cocycle $(\Us_t)_t$ is the one-parameter group of unitaries
defined by
\be
\Us_t&=&e^{istH},\qquad t\in\Ro.
\nnee
The analytic continuation of $t\mapsto \Us_t\Omega_0=e^{istH}\psi$ is
\be
\Xi^{(s)}_z&=&e^{iszH}\psi.
\nnee
The vector $\Xi^{(s)}_z$ belongs to the Hilbert space $\cal H$ because
$|s\Im z|<1/2$ implies that
\be
\int\upd x\,|\Xi^{(s)}_z(x)|^2
&=&\int\upd x\,e^{-2s(\Im z)h(x)}\,|\psi(x)|^2\cr
&\le&\int\upd x\,e^{-|h(x)|}\,|\psi(x)|^2\cr
&<&+\infty.
\nnee
The latter inequality holds because of item 1) of the Proposition.

The vector $\Omega_s$ is given by 
\be
\Omega_s=e^{-\frac 12\zeta(s)}\Xi^{(s)}_{i/2}=e^{-\frac 12\zeta(s)}e^{-\frac 12sH}\psi,
\nnee
with
\be
\zeta(s)&=&\log \int\upd x\,e^{-sh(x)}|\psi(x)|^2.
\nnee
The operator $T_s$ is given by
\be
T_s=\Usi=e^{-\frac 12sH}.
\nnee
It satisfies $\log T_s=-\frac 12sH$. Therefore, the requirement for the geodesic
to be log-affine is fufilled with a vanishing function $\Phi(s)$.

\endproof

Finally, note that in this abelian case the operators $X_s$ and $T_s$ coincide.

\section{Summary}

Section \ref {subsect:def} proposes a definition for the notion of a log-affine geodesic
$s\in[0,1]\mapsto\omega_s$
connecting two states of a von Neumann algebra ${\cal A}$.
With each state $\omega_s$ corresponds a cocycle $\Us$ of the modular automorphism group
of the starting point $\omega_0$. The geodesic is said to be log-affine when the 
positive operator $T_s$ associated with the cocycle $\Us$ is a log-affine function of $s$,
up to a scalar normalization factor.
The definition is linked to that of existing notions
in the case of a finite-dimensional Hilbert space and in the abelian case.
The definition generalizes the notion of a trajectory which starts at a tracial state
and which is described by the action of a semigroup.

The definition of log-affine geodesics as given here is only a first step in developing a general theory
of the geometry of manifolds consisting of vector states on a von Neumann algebra
and belonging to an exponential family.

\appendix

\section{Appendix}

Here, items 5) and 6) of Proposition \ref {geo:prop4} are proved.
The essence of the proof is that $U_t\Delta^{-it}=T^{-it}$,
where $T$ is a positive operator.
In a finite-dimensional Hilbert space one has 
\be
\Ui\Delta^{1/2}=T^{1/2}=\left(T^{1/2}\right)^*=\Delta^{1/2}\left(\Ui\right)^*.
\nnee
This implies
\be
J\Xi_{i/2}=J\Ui\Omega=J\Ui\Delta^{1/2}\Omega=J\Delta^{1/2}\left(\Ui\right)^*\Omega=S\left(\Ui\right)^*\Omega
=\Ui\Omega=\Xi_{i/2}
\nnee
and
\be
Y=J\Ui J=J\Delta^{1/2}\left(\Ui\right)^*\Delta^{-1/2}J
=JT^{1/2}S.
\nnee
In the general case one has to take care of domain problems.

\begin{proposition}
\label{app:prop}
 For all $A$ in $\cal A$ is $A\Omega_0$ in the domain of $T^{1/2}$ and of $\Ui\Delta^{1/2}$
 and one has
 \be
 T^{1/2}A\Omega_0=U_{i/2}\Delta^{1/2}A\Omega_0=JA^*J\Xi_{i/2}.
 \nnee
\end{proposition}

\beginproof

The vector $JA^*J\Omega_0=JA^*\Omega_0=\Delta^{1/2}A\Omega_0$ belongs to the domain of $\Delta^{-1/2}$
and also to the domain of $\Ui$, with $\Ui JA^*J\Omega_0=JA^*J\Ui\Omega_0=JA^*J\Xi_{i/2}$.
Therefore $A\Omega_0$ belongs to the domain of $\Ui\Delta^{1/2}$
and one has $\Ui\Delta^{1/2}A\Omega_0=JA^*J\Xi_{i/2}$.

Next take any $\psi$ in the domain of $T^{1/2}$.
Then $(T^{1/2}\psi,A\Omega_0)$ is the analytic continuation of $(T^{-it}\psi,A\Omega_0)$
to the point $t=i/2$.
One has
\be
(T^{-it}\psi,A\Omega_0)
&=&
(U_t\Delta^{it}\psi,A\Omega_0)\cr
&=&
(\psi,U_{-t}\Delta^{it}A\Omega_0).
\nnee
Take $t=i/2$  to obtain
\be
(T^{1/2}\psi,A\Omega_0)&=&(\psi,U_{i/2}\Delta^{1/2}A\Omega_0).
\nnee
This shows that $A\Omega_0$ belongs to the domain of $T^{1/2}$ and that
$T^{1/2}A\Omega_0=U_{i/2}\Delta^{1/2}A\Omega_0$.

\endproof

\paragraph{Proof of item 5)}
The above Proposition is used in the following calculation.
For all $Z$ in the commutant ${\cal A}'$ one has
\be
(Z\Omega_0,\Xi_{i/2})
&=&(\Omega_0,Z^*\Xi_{i/2})\cr
&=&(\Omega_0,T^{1/2}JZ\Omega_0)\cr
&=&(T^{1/2}\Omega_0,JZ\Omega_0)\cr
&=&(Z\Omega_0,JT^{1/2}\Omega_0).
\nnee
This shows that $\Xi_{i/2}=JT^{1/2}\Omega_0=J\Xi_{i/2}$.

\paragraph{Proof of item 6)}

Note that $\Omega_0$ is in the domain of $Y=J\Ui^{**}J$
and that $Y\Omega_0=J\Ui \Omega_0=J\Xi_{i/2}=\Xi_{i/2}$.
This implies that $\Omega_0$ is also in the domain of $X^{1/2}$
and that $X^{1/2}\Omega_0=K^*T^{1/2}\Omega_0=K^*\Xi_{i/2}$,
with $Y=KX^{1/2}$ the polar decomposition of $Y$.
From Proposition \ref{app:prop} it follows that for all $A\in{\cal A}$
\be
T^{1/2}SA\Omega_0=JAJ\Xi_{i/2}=JAY\Omega_0=JYA\Omega_0.
\nnee
Because ${\cal A}\Omega_0$ is a core of $Y$ this implies that $JT^{1/2}S$ is
an extension of $Y$.

\section*{}

\end{document}